\documentclass[11pt,a4paper]{article}
\usepackage{graphicx}  
\usepackage{dcolumn}   
\usepackage{bm}        
\usepackage{amssymb}   
\usepackage[latin1]{inputenc}
\usepackage{amsmath}
\usepackage{amsfonts}
\usepackage{amssymb}
\usepackage{setspace}
\usepackage{mathtools}
\usepackage{pstricks}
\usepackage{color}
\usepackage{bbold}
\usepackage{jheppub}
\usepackage{ulem}

\providecommand{\f}[2]{\frac{{#1}}{{#2}}}

\newcommand{\ee}[1]{\begin{equation}#1\end{equation}}
\newcommand{\ea}[1]{\begin{align}#1\end{align}}

\def\undertilde#1{\mathord{\vtop{\ialign{##\crcr
$\hfil\displaystyle{#1}\hfil$\crcr\noalign{\kern1.5pt\nointerlineskip}
$\hfil\tilde{}\hfil$\crcr\noalign{\kern1.5pt}}}}}

\title{Massive scalar field evolution in de Sitter}

\author[a]{Tommi Markkanen}
\author[b]{and Arttu Rajantie}
\affiliation[a]{Department of Physics, King's College London, Strand, London WC2R 2LS, UK}
\affiliation[b]{Department of Physics, Imperial College London, SW7 2AZ, UK}

\abstract{The behaviour of a massive, non-interacting and non-minimally coupled quantised scalar field
in an expanding de Sitter background is investigated by solving the field evolution for an arbitrary initial state. In this approach there is no need to choose a vacuum in order to provide a definition for particle states, nor to introduce an explicit ultraviolet regularization. 
We conclude that the expanding de Sitter space is a stable equilibrium configuration under small perturbations of the initial conditions. Depending on the initial state, the energy density can approach its asymptotic value from above or below, the latter of which implies a violation of the weak energy condition. The backreaction of the quantum corrections can therefore lead to a phase of super-acceleration also in the non-interacting massive case.}

\emailAdd{tommi.markkanen@kcl.ac.uk}
\emailAdd{a.rajantie@imperial.ac.uk}

\begin{document}

\maketitle

\section{Introduction}

An eternally expanding de Sitter space has been shown in the classical context to be a stable attractor solution for a variety of cosmological models with a cosmological constant in the context of cosmic no-hair theorems \cite{Wald:1983ky,Starobinsky:1982mr,Gibbons:1977mu,Hawking:1981fz} (see \cite{Schmidt:2004xv,Brandenberger:2016uzh} for review and references). For quantized theories de Sitter space has been analyzed in terms of back-reaction of cosmological fluctuations in \cite{Tsamis:1996qq2,Tsamis:1996qq3,Geshnizjani:2002wp} quantum gravity in \cite{Tsamis:1996qq,Tsamis:1996qq0,Tsamis:1996qq1,Antoniadis:1985pj} and quantum fields on a classical curved background
in \cite{ford,Bousso:2001mw,Polyakov:2007mm,Polyakov:2007mm1, Polyakov:2007mm2,akh,Marolf:2010zp,Marolf:2010nz,Mottola:1984ar,Anderson:2013ila,Anderson:2013ila1,Habib:1999cs, Anderson:2013ila2,Onemli:2004mb,Onemli:2002hr,Dabrowski:2014ica,Antoniadis:2006wq,Anderson:2000wx,Koivisto:2010pj, Albrecht:2014hxa,Glavan:2014uga,Glavan:2015cut,Jiang:2016nok}. In \cite{Anderson:2013ila,Anderson:2013ila1} a thorough investigation of de Sitter space in a global coordinate system was presented and as the main finding it was stated that the global de Sitter space with a quantized scalar field is unstable thus reinforcing the arguments made already in \cite{Mottola:1984ar}. 

Quite generically a quantum field coupled to a curved background leads to the creation of quanta which is often called cosmological particle production \cite{Parker:1968mv,Parker:1968mv1,Parker:1968mv2,Parker:1968mv3}, or in the case of a black hole, Hawking radiation  \cite{Hawking:1974sw,Hawking:1974rv}. Also in flat space there are closely related processes: in Schwinger pair creation particles are produced due to a background electric field \cite{Schwinger:1951nm} (for a recent review, see \cite{Gelis:2015kya}) and in the Unruh effect because of constant acceleration \cite{Unruh:1976db}. A crucial conclusion from Hawking radiation is that black holes may be viewed as thermodynamic objects, which is manifested as the relation between the event horizon area and entropy of a black hole \cite{Bekenstein0,Bekenstein1,Bekenstein2,Bardeen:1973gs,Bekenstein3,Hawking:1976de}.
Importantly,  analogous thermal properties have also been assigned space-times with boundaries \cite{Gibbons:1977mu}, which can be assigned entropy and temperature, a connection made even more profound by the observation that in many cases it is possible to derive the Einstein equations from thermodynamics \cite{Jacobson:1995ab,Padmanabhan:2002sha,kof,cai,Padmanabhan:2009vy}. However, it is not obvious how far one may take the thermodynamic analogy of de Sitter space as discussed in \cite{Davies:2003me,Birrell:1982ix}.

Based on the purely thermodynamic description of \cite{Gibbons:1977mu}, de Sitter space has been investigated for example in \cite{Antoniadis:2006wq,Mottola:1985qt,Sekiwa:2006qj,Padmanabhan:2003gd}.
The thermal features in de Sitter space are often studied in the static coordinates with explicitly a horizon and a coordinate singularity much like in the Schwarzschild metric. In such coordinates de Sitter space appears then essentially as an insulated cavity that is internally in equilibrium containing a thermal spectrum of particles \cite{kof,Kaloper:2002cs,Susskind:2003kw}. However, our current view of the Universe is best described as that of a co-moving observer in the flat FLRW coordinates
$ds^2=-dt^2+a(t)^2d\mathbf{x}^2$, which warrants a detailed investigation of de Sitter space in this specific coordinate system.

In this work we set out to investigate the evolution of a quantized scalar field in expanding de Sitter space and its implications for the backreaction in the flat FLRW coordinates. We study the backreaction in the semi-classical approach with a strictly classical geometry. Our model consists of a quantized, massive, noninteracting and non-minimally coupled scalar field in a classical spacetime curved by vacuum energy. 

One of our key findings is that quantum energy-momentum tensor  of a non-interacting massive scalar field
can violate the weak energy condition. A similar result was found earlier 
for a massless, minimally coupled and self-interacting model in \cite{Onemli:2004mb,Onemli:2002hr}. 
In that case the origin of such a violation was the generation of a nonperturbative mass via interactions, as described in \cite{Starobinsky:1994bd}, see also the related discussion in \cite{Winitzki:2001fc}. A violation of the weak energy condition in the massive and non-interacting model has
different origins and highlights the generality of the effect.

Furthermore, for our study we define a new covariant renormalization prescription for de Sitter space. Our method has all the benefits of the popular adiabatic subtraction technique \cite{Parker:1974qw,Parker:1974qw1,Bunch:1980vc}, but it does not require the lengthy expressions needed in adiabatic subtraction and is physically more meaningful.

Our units are $\hbar\equiv c\equiv k_B\equiv1$ and we use the (+,+,+) conventions of \cite{Misner:1974qy}.

\section{Modes in de Sitter space}
\label{sec:br}

The FLRW coordinates for flat de Sitter space with Hubble parameter $H$ are $ds^2=-dt^2+a(t)^2d\mathbf{x}^2$, where the scale factor is written as $a(t)=e^{Ht}\equiv a$. The matter Lagrangian for the massive non-minimally coupled scalar field reads in $n$ spacetime dimensions
\ee{S_m=-\int d^nx\sqrt{-g}\,\left[\f{1}{2}(\nabla_\mu \phi)^2+\f{1}{2}m^2\phi^2+\f{1}{2}\xi R\phi^2\right]\, ,\label{eq:act1}} and the scalar field $\hat\phi$ satisfies the equation of motion
\ee{\left(-\Box+m^2+\xi R\right)\hat{\phi}=0\,, \label{eq:eom}}
where $R=n(n-1)H^2$. The solutions to (\ref{eq:eom}) can be expressed via modes, which for generality we also write in $n$ dimensions
\ee{\hat{\phi}=\int \f{d^{n-1}\mathbf{k}\, e^{i\mathbf{k\cdot\mathbf{x}}}}{\sqrt{(2\pi a)^{n-1}}}\left[\hat{a}_\mathbf{k}^{\phantom{\dagger}}f^{\phantom{\dagger}}_\mathbf{k}(t)+\hat{a}_{-\mathbf{k}}^\dagger f^*_{-\mathbf{k}}(t)\right]\,\label{eq:adsol2}\, ,}
with $[\hat{a}_{\mathbf{k}}^{\phantom{\dagger}},\hat{a}_{\mathbf{k}'}^{\phantom{\dagger}}]=[\hat{a}_{\mathbf{k}}^{{\dagger}},\hat{a}_{\mathbf{k}'}^\dagger]=0$ and $[\hat{a}_{\mathbf{k}}^{\phantom{\dagger}},\hat{a}_{\mathbf{k}'}^\dagger]=\delta^{(n-1)}(\mathbf{k}-\mathbf{k}')$, 
where $\mathbf{k}$ denotes the comoving momentum.
The mode equation from (\ref{eq:eom}) with (\ref{eq:adsol2}) is
\ee{\ddot{f}_{\mathbf{k}}(t)+\omega^2_{\mathbf{k}}{f}_\mathbf{k}(t)=0\,,\label{eq:modds}} with the definitions \ee{\omega_{\mathbf{k}}^2=\f{\mathbf{k}^2}{a^2}+{H^2}\gamma^2\,,\qquad \gamma\equiv\bigg[\f{m^2}{H^2}-\f{(n-1)^2}{4}+n(n-1)\xi\bigg]^{1/2}\approx \f{m}{H} \,,\label{equ:defomega}}
where the last approximation follows if $m\gg H$ and $\xi\sim 1$. As we are interested in the massive case we assume $\gamma\gtrsim1$ throughout.
As discussed in \cite{Anderson:2013ila,Anderson:2013ila1}, in de Sitter the general mode solutions of the mode equation (\ref{eq:modds}) have two special classes, which are not orthogonal, the ``in'' and the ``out'' modes
\ea{
f^{\rm in}_\mathbf{k}(t)
&=
\sqrt{\f{\pi}{4H}}e^{-\pi\gamma/2}H^{(1)}_{i\gamma}\left(\f{\vert\mathbf{k}\vert}{a H}\right),\label{eq:hank}\
\\ f^{\rm out}_\mathbf{k}(t)&=\left(\f{2H}{\vert\mathbf{k}\vert}\right)^{i\gamma}\f{\Gamma[1+i\gamma]}{\sqrt{2H\gamma}}J_{i\gamma}\left(\f{\vert\mathbf{k}\vert}{a H}\right),
 \label{eq:bes}}
where $H^{(1)}_{i\gamma}(z)$ and $J_{i\gamma}(z)$ are the Hankel and Bessel functions of the first kind respectively and have the asymptotic relations \cite{AW}
\ea{H_{i\gamma}^{(1)}(z) &=\sqrt{\frac{2}{\pi z}}\exp\bigg\{i\bigg(z-\frac{i\gamma\pi}{2}-\frac{\pi}{4}\bigg)\bigg\}\,,\quad z \gg \vert\gamma^2 +1/4\vert\,;\label{eq:asy1}\\ J_{i\gamma}(z) &=\frac{1}{\Gamma[1+i\gamma]} \left( \frac{z}{2} \right) ^{i\gamma}\,,\qquad\qquad\quad~\,\quad z^2\ll \vert i\gamma+1\vert\,.\label{eq:asy2}} At early times $t\rightarrow-\infty$ or large momenta, the physical wavenumber of the mode \ee{k_{\rm ph}\equiv |\mathbf{k}|/a\,,} is much larger than the Hubble rate $H$. We will impose the requirement that the short distance behaviour of the mode coincides with that in flat Minkowski space \cite{Allen:1985ux}. More precisely, when the frequency $\omega_{\mathbf{k}}$ given by (\ref{equ:defomega}) satisfies the adiabaticity condition $|\dot\omega_{\mathbf{k}}|\ll \omega_{\mathbf{k}}^2$, we can write an approximate solution to (\ref{eq:modds})
\ee{
f^{\rm ad}_{\mathbf{k}}(t)=\f{\exp\left\{-i\int^t \omega_{\mathbf{k}}\right\}}{\sqrt{2 \omega_{\mathbf{k}}}}\,.
\label{equ:fad}
}
We can clearly see (\ref{equ:fad}) to contain only the positive frequency component and coincide with the usual Minkowski vacuum mode when $a\equiv1$. The mode in (\ref{equ:fad}) is generally called the adiabatic mode, which in curved space in many instances is the closest to the usual definition of vacuum in Minkowski space \cite{Birrell:1982ix,Parker:2009uva}.

The ``in'' solution (\ref{eq:asy1}) can be recognized as the Bunch-Davies vacuum \cite{Chernikov:1968zm,BD}. 
Its asymptotic behavior as $k_{\rm ph}\rightarrow \infty$ coincides with the adiabatic mode (\ref{equ:fad}), which implies that it is the solution that describes a mode that was in the vacuum at early times. It also implies that all modes with very large momenta behave approximately as the Minkowski vacuum mode. Similarly for  $f^{\rm out}_\mathbf{k}$ we can show from (\ref{eq:asy2}) that when  $k_{\rm ph}\rightarrow 0$ it coincides with (\ref{equ:fad})

The behaviour of the solution $f^{\rm in}_\mathbf{k}$ with respect to time can be interpreted as particle production. To see this, first we note that the mode expansions (\ref{eq:hank}) and (\ref{eq:bes}) can be connected via a Bogolubov transformation \cite{Anderson:2013ila,Anderson:2013ila1}
\ee{f^{\rm in}_\mathbf{k}(t)=\alpha^{\phantom{\dagger}}_{\mathbf{k}}{f}^{\rm out}_{\mathbf{k}}(t)+\beta^{\phantom{\dagger}}_{\mathbf{k}}{f}^{\rm out \,*}_{\mathbf{k}}(t) \,;\qquad f^{\rm out}_\mathbf{k}(t)=\alpha^{*}_{\mathbf{k}}{f}^{\rm in}_{\mathbf{k}}(t)-\beta^{\phantom{\dagger}}_{\mathbf{k}}{f}^{\rm in\,*}_{\mathbf{k}}(t)\label{eq:BEbo}} with \ee{\beta_\mathbf{k}=\left(\f{2H}{|\mathbf{k}|}\right)^{i\gamma}\f{e^{\pi\gamma/2}\sqrt{2\pi\gamma}}{(1-e^{2\pi\gamma})\Gamma[1-i\gamma]}\,; \qquad\alpha_\mathbf{k}=-e^{\pi\gamma}\beta_\mathbf{k}^*\quad\Rightarrow \quad\vert\beta_\mathbf{k}\vert^2=\f{1}{e^{2\pi \gamma}-1}.\label{eq:B-E}}
Since at very early times $f^{\rm in}_\mathbf{k}$ coincides with (\ref{equ:fad}) and similarly for $f^{\rm out}_\mathbf{k}$ for late times, from (\ref{eq:BEbo}) we can schematically write the evolution for the ``in'' mode between $t\rightarrow\pm\infty$ \ee{
f^{\rm in}_\mathbf{k}(-\infty)=f^{\rm ad}_\mathbf{k}(-\infty)\quad\longrightarrow\quad f^{\rm in}_\mathbf{k}(\infty)=\alpha_\mathbf{k}f^{\rm ad}_\mathbf{k}(\infty)
+\beta_\mathbf{k}f^{\rm ad*}_\mathbf{k}(\infty).
\label{eq:split}} 
Since the mode $f^{\rm ad}_\mathbf{k}$ represents a vacuum solution this shows that $\vert\beta_\mathbf{k}\vert^2$ can be interpreted as the number of particles created from the vacuum. We can also illustrate this process by using the adiabatic invariant \cite{Kofman:1997yn}
\ee{n_{\mathbf{k}}(t)
\equiv \f{\omega_{\mathbf{k}}}{2}\bigg(\f{\vert\dot{f}_\mathbf{k}(t)\vert^2}{\omega_{\mathbf{k}}^2}+\vert{f}_\mathbf{k}(t)\vert^2\bigg)-\f{1}{2}\,,\label{eq:KLS}}
which to some extent can be interpreted as the particle number. From (\ref{eq:BEbo}) we can find that for the in-mode
\ea{
n_\mathbf{k}(t)&\xrightarrow{t\rightarrow-\infty} 0,\nonumber\\
n_\mathbf{k}(t)&\xrightarrow{t\rightarrow\infty} \vert\beta_\mathbf{k}\vert^2, }
and that for the out-mode
\ea{
n_\mathbf{k}(t)&\xrightarrow{t\rightarrow-\infty} \vert\beta_\mathbf{k}\vert^2,\nonumber\\
n_\mathbf{k}(t)&\xrightarrow{t\rightarrow\infty} 0. }
By studying the large and small argument expansions of the Hankel and Bessel functions from (\ref{eq:asy1}) and (\ref{eq:asy2}) for $m\gg H$ we can roughly approximate that the branching of the single positive frequency mode to a linear combination of positive and negative frequencies as in (\ref{eq:split}) takes place when the physical momentum of the mode satisfies
\ee{\sqrt{mH}\lesssim k_{\rm ph}\lesssim m^2/H
\,.\label{eq:IM}}
\begin{figure}
\begin{minipage}{1\textwidth}
\center
\hspace{1.5cm}
\includegraphics[width=11cm]{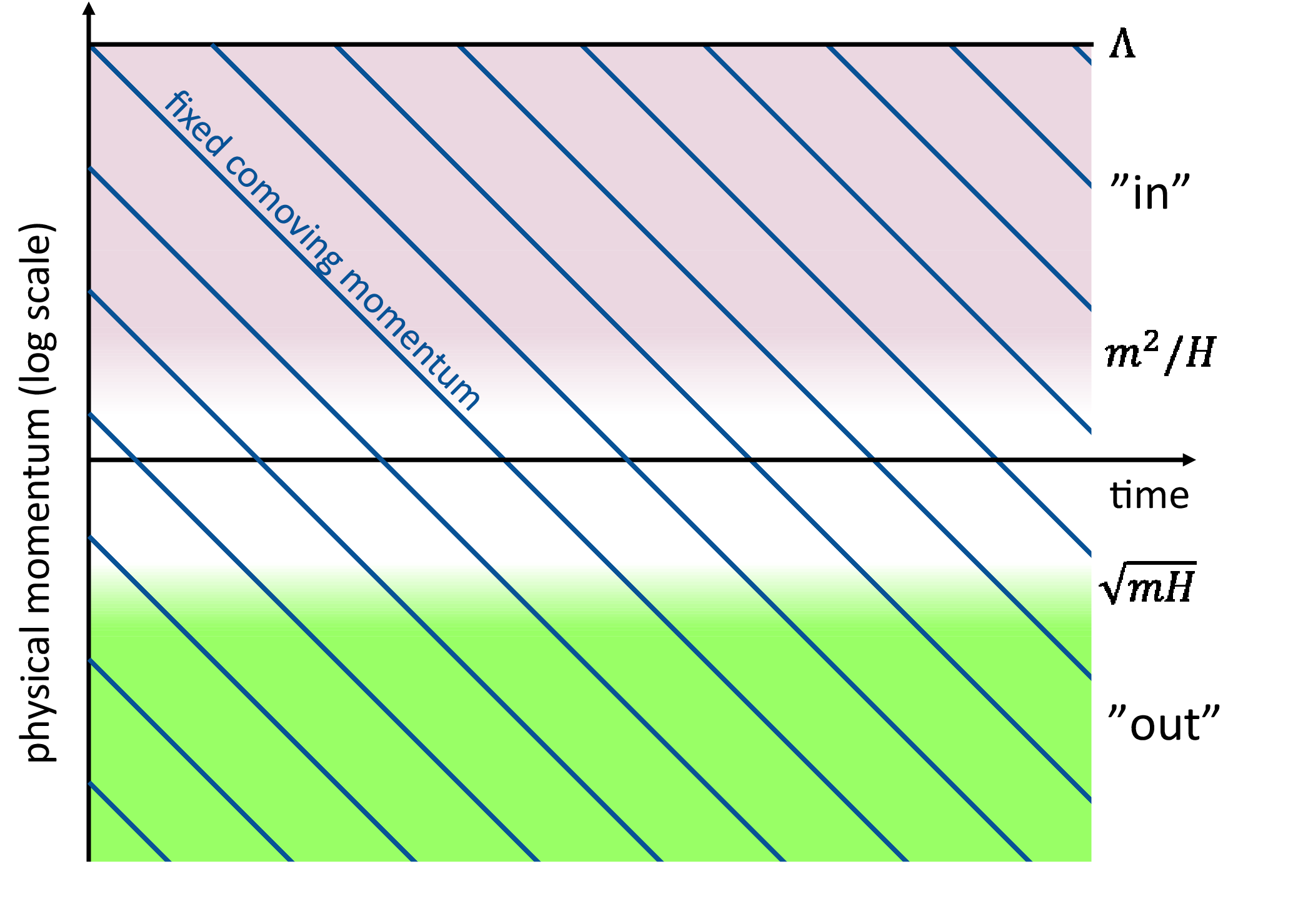}
\end{minipage}
\caption{Illustration of the time evolution of comoving modes in terms of physical momentum. The diagonal lines show the physical momentum of individual comoving mode. In the coloured ``in'' and ``out'' regions the adiabatic mode (\ref{equ:fad}) is well approximated by Eqs.~(\ref{eq:hank}) and (\ref{eq:bes}), respectively. 
A comoving mode that starts in the ``in'' region experiences non-trivial time evolution as it passes through the white ``particle production'' band, but in terms of the physical momentum the situation is stationary.
If we impose an ultraviolet cutoff at fixed physical momentum $\Lambda$, new comoving modes emerge from the cutoff at a constant rate.
\label{fig1}
} 
\end{figure}
\begin{figure}
\begin{minipage}{1\textwidth}
\center
\hspace{-1.5cm}
\includegraphics[width=0.78\textwidth,trim={0cm 5cm  0cm  8cm },clip]{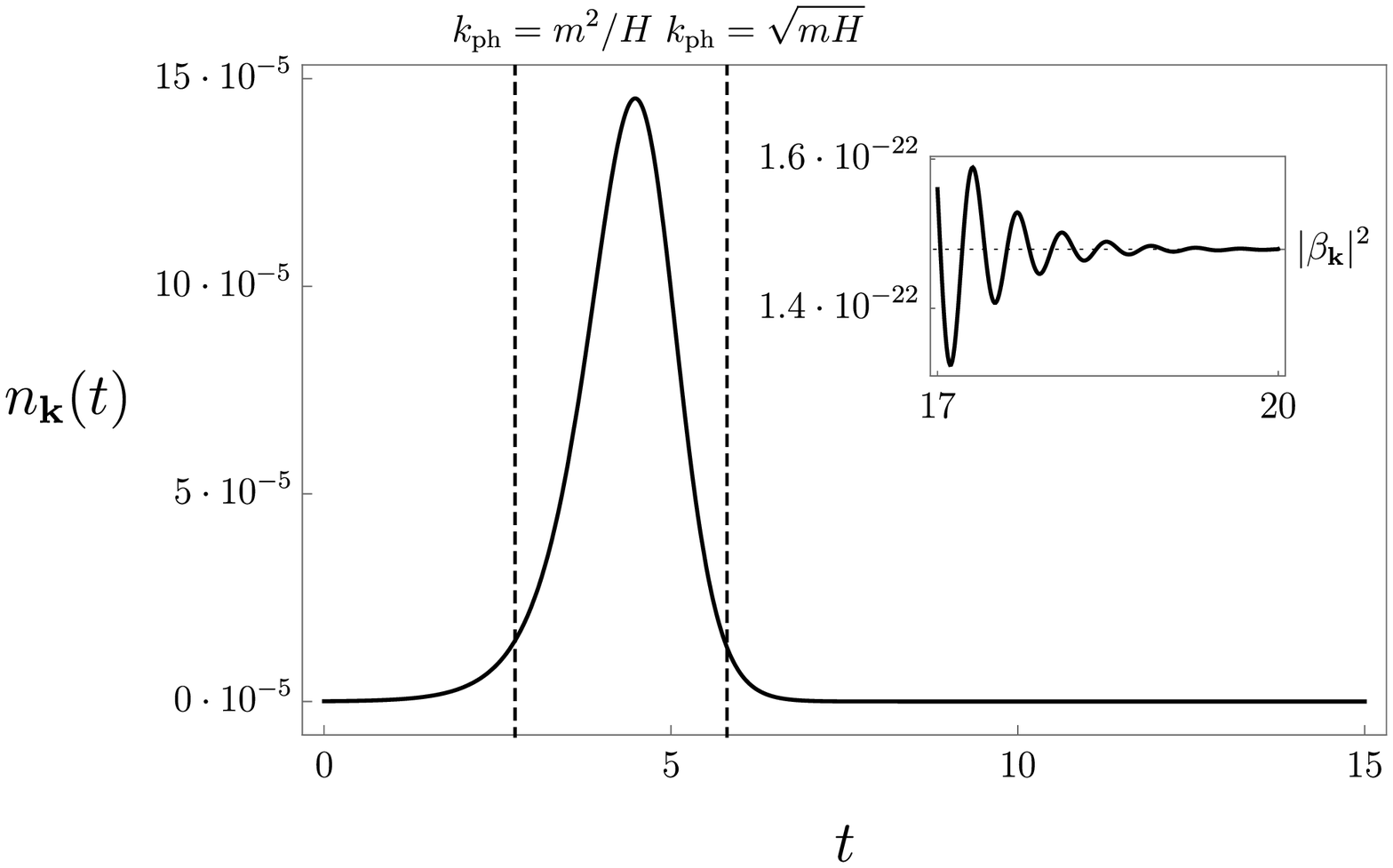}
\end{minipage}
\vspace{-2cm}
\caption{Behaviour of (\ref{eq:KLS}) for the ``in'' mode with $\vert\mathbf{k}\vert=10^8$ and $\gamma=8$ with respect to time in the units of $H=1$. The region (\ref{eq:IM}) is given by the dashed lines. In the inset we show that the very late time behaviour approaches $\vert\beta_\mathbf{k}\vert^2\sim e^{-2\pi\gamma}\sim1.48\cdot10^{-22}$ despite the particle number being significantly larger at intermediate times.   \label{fig2} 
} \end{figure}
\!\!In Fig.\ref{fig1}\, we present a schematic illustration of the behaviour of the quantum modes as a function of time. The important feature is that constantly more and more new modes are redshifted from the ultraviolet (UV) towards the infrared (IR), which essentially scatter from the curvature as (\ref{eq:split}) once the they exit the region (\ref{eq:IM}). In Fig.\ref{fig2}\, we shown the behaviour of (\ref{eq:KLS}) for the ``in'' mode starting with large physical momentum. What we can there see is that even though the final particle number $\vert\beta_\mathbf{k}\vert^2$ is exponentially suppressed, as given by (\ref{eq:B-E}), it shows a large fluctuation at intermediate times. 

It is easy to see that a Bose-Einstein distribution with a small Gibbons-Hawking temperature $T_{\rm dS}=H/(2\pi)\ll m$ \cite{Gibbons:1977mu}, approaches the $\vert\beta_\mathbf{k}\vert^2$ term in (\ref{eq:B-E}) when $H\rightarrow 0$, as for a small temperature the average momentum scales as $\langle k^2\rangle\sim mH$ so one can write $\omega_{\mathbf{k}}/T_{\rm ds}\sim 2\pi m/H\sim 2\pi \gamma$.

On the other hand, by expressing Eq.~(\ref{eq:hank}) in terms of the physical momentum $\mathbf{k}_{\rm ph}=\mathbf{k}/a$, one can see that the ``in'' solutions correspond to an explicitly time-independent equilibrium state of the field, 
\begin{equation}
\label{equ:feq}
f^{\rm in}_{a\mathbf{k}_{\rm ph}}=f^{\rm eq}({k}_{\rm ph})\equiv
\sqrt{\f{\pi}{4H}}e^{-\pi\gamma/2}H^{(1)}_{i\gamma}\left(\f{{k}_{\rm ph}}{H}\right),
\end{equation}
which is just the Bunch-Davies vacuum state.
The time derivative $\dot{f}^{\rm in}_{\mathbf{k}}(t)$ is then given by
\ee{\dot{f}^{\rm in}_{a\mathbf{k}_{\rm ph}}=-H k_{\rm ph}\f{\partial}{\partial k_{\rm ph}}{f}^{\rm eq}(k_{\rm ph})\,,\label{equ:feq1}}
and similarly for higher derivatives. 
\section{The Energy-Momentum Tensor}
\label{sec:EM}
As discussed, the time evolution of a quantum mode in de Sitter can be interpreted as particle creation. In this work however our main focus will not be on the generation of particles, as it is well-known that particle number in curved space is an observer-dependent quantity  leading to ambiguities in terms of its proper definition \cite{Davies:2003me}. Instead we will focus on the energy-momentum tensor, for which one is not compelled to make any reference to particles. All one needs to define is an initial condition and a renormalization prescription. As we will show, the behaviour of the energy-momentum tensor at equilibrium one may relatively easily derive from first principles. The complete energy-momentum is
\ee{
T_{\mu}^{~\nu}\equiv\mbox{diag}(\rho(t)+\rho_{\rm vac},p(t)+p_{\rm vac},p(t)+p_{\rm vac},p(t)+p_{\rm vac})\,,
\label{equ:Tmunu0}}
where $\rho_{\rm vac}=-p_{\rm vac}$ is the contribution from the vacuum energy, and where the $\rho(t)$ and $p(t)$ are the contributions from the quantum field, whose energy momentum tensor reads \cite{Markkanen:2013nwa} 
\ee{\hat{T}_{\mu\nu}^\phi=-\f{g_{\mu\nu}}{2}\big[\partial_\rho\hat{\phi}\partial^\rho\hat{\phi}+m^2\hat{\phi}^2\big]+\partial_\mu\hat{\phi}\partial_\nu\hat{\phi} +\xi\big[G_{\mu\nu}-\nabla_\mu\nabla_\nu+g_{\mu\nu}\Box\big]\hat{\phi}^2\, .\label{eq:em}}
Its expectation value defines the energy and pressure densities $\rho$ and $p$ as 
\ee{
\mbox{diag}\left(\rho(t),p(t),p(t),p(t)\right)
\equiv\langle\hat{T}_{\mu}^{\phi\,\nu}\rangle
-\mbox{diag}\left(\delta\rho,\delta p,\delta p,\delta p\right)\,,
\label{eq:emr}}
where we have introduced the renormalisation counterterms $\delta\rho$ and $\delta p$, which reflect the freedom in splitting the overall energy-momentum tensor between the field and the vacuum in Eq.~(\ref{equ:Tmunu0}). As always in renormalisation, the choice of the counterterms is arbitrary and does not affect physics because any change in them can be absorbed by changing $\rho_{\rm vac}$ and $p_{\rm vac}$

In momentum space we can write
\ea{
\rho(t)=\int \f{d^{n-1}\mathbf{k}\, }{{(2\pi a)^{n-1}}}\rho_{\mathbf{k}}(t)-\delta\rho,\quad
p(t)=\int \f{d^{n-1}\mathbf{k}\, }{{(2\pi a)^{n-1}}}p_{\mathbf{k}}(t)-\delta p,\label{equ:modecontrib}}
where $\rho_\mathbf{k}(t)$ and $p_\mathbf{k}(t)$ are the contributions from the comoving mode $\mathbf{k}$ to the energy density and pressure, respectively.
However, because the integrals are ultraviolet divergent, one needs to proceed with care.

Using Ref.~\cite{Markkanen:2013nwa} we can write them as
\ea{
\rho_{\mathbf{k}}(t) &=\f{1}{2}\Big\vert \dot{f}^{}_{\mathbf{k}}(t)-\f{(n-1)H}{2}{f}^{}_{\mathbf{k}}(t)
\Big\vert^2\nonumber \\&+\f{1}{2}\bigg[\f{\mathbf{k}^2}{a^2}+m^2-2\xi\bigg(\f{n(n-1)}{2}H^2-(n-1)H\partial_t\bigg)\bigg]\big\vert {f}^{}_{\mathbf{k}}(t)\big\vert^2\,,\label{eq:ED1}}
and 
\ea{
p_{\mathbf{k}}(t) &=\f{1}{2}\Big\vert \dot{f}^{}_{\mathbf{k}}(t)-\f{(n-1)H}{2}{f}^{}_{\mathbf{k}}(t)
\Big\vert^2\nonumber \\&-\f{1}{2}\bigg[\f{(n-3)\mathbf{k}^2}{(n-1)a^2}+m^2+2\xi\bigg(\f{n(n-1)}{2}H^2-nH\partial_t+\partial_t^2\bigg)\bigg]\big\vert {f}^{}_{\mathbf{k}}(t)\big\vert^2\,.\label{eq:PD1}}

It follows directly from the equation of motion (\ref{eq:modds}) that the mode contributions satisfy
\ee{\frac{d\rho_{\mathbf{k}}(t) }{dt}+(n-1)Hp_{\mathbf{k}}(t)=0.\label{equ:modeencons}}
If the integrals (\ref{equ:modecontrib}) were convergent and we could ignore the counterterms, this would imply that $\rho$ and $p$ satisfy the continuity equation, $\nabla^\mu T_{\mu\nu}=0$ or
\ee{\frac{d\rho(t)}{dt}+(n-1)H\big(p\left(t\right)+\rho\left(t\right)\big)=0\,,\label{equ:modeencons2}}
which would lead to the conclusion that if a configuration has a time independent energy density, it must satisfy 
\ee{p(t)+\rho(t)=0\,,\label{equ:modeencons3}}
in de Sitter space. However, the integrals are actually ultraviolet divergent, leading to some subtleties, which we will next discuss.

For this purpose, it is useful to express the energy density and pressure in terms of contributions 
$\tilde{\rho}_{\mathbf{k}_{\rm ph}}$, 
$\tilde{\rho}_{\mathbf{k}_{\rm ph}}$ as functions of the physical modes $\mathbf{k}_{\rm ph}=\mathbf{k}/a$
\ea{
\rho(t)=\int \f{d^{n-1}\mathbf{k}_{\rm ph}\, }{{(2\pi)^{n-1}}}\tilde\rho_{\mathbf{k}_{\rm ph}}(t)-\delta \rho,\quad
p(t)=\int \f{d^{n-1}\mathbf{k}_{\rm ph}\, }{{(2\pi)^{n-1}}}\tilde{p}_{\mathbf{k}_{\rm ph}}(t)-\delta{p},\label{equ:phmodecontrib}}
where the tilde $~\tilde{ }~$ indicates a contribution from a physical rather than a comoving mode.
The two sets of quantities are related by
\begin{equation}
\tilde{\rho}_{\mathbf{k}_{\rm ph}}(t)=\rho_{a\mathbf{k}_{\rm ph}}(t),\quad
\tilde{p}_{\mathbf{k}_{\rm ph}}(t)=p_{a\mathbf{k}_{\rm ph}}(t).
\end{equation}

In the Bunch-Davies equilibrium state (\ref{equ:feq}), with the help of (\ref{equ:feq1})  the physical mode contributions to the energy density and pressure are time indepedent and given by Eqs.~(\ref{eq:ED1}) and (\ref{eq:PD1}), calculated using the ``in'' mode,
\ea{
\tilde{\rho}^{\rm eq}({k}_{\rm ph})&=\f{H^2}{2}\Big\vert k_{\rm ph}\f{\partial {f}^{\rm eq}(k_{\rm ph})}{\partial k_{\rm ph}}+\f{(n-1)}{2}{f}^{\rm eq}(k_{\rm ph})
\Big\vert^2\nonumber \\&+\f{1}{2}\bigg[k_{\rm ph}^2+m^2-2\xi\bigg(\f{n(n-1)}{2}H^2+(n-1)H^2 k_{\rm ph}\f{\partial}{\partial k_{\rm ph}}\bigg)\bigg]\big\vert {f}^{\rm eq}(k_{\rm ph})\big\vert^2\,,\label{eq:EDeq}}
and 
\ea{
\tilde{p}^{\rm eq}({k}_{\rm ph}) &=\f{H^2}{2}\Big\vert  k_{\rm ph}\f{\partial {f}^{\rm eq}(k_{\rm ph})}{\partial k_{\rm ph}}+\f{(n-1)}{2}{f}^{\rm eq}(k_{\rm ph})
\Big\vert^2\nonumber \\&-\f{1}{2}\bigg[\f{(n-3)}{(n-1)}{k}_{\rm ph}^2+m^2\nonumber \\&+2\xi\bigg(\f{n(n-1)}{2}H^2+(n+1)H^2k_{\rm ph}\f{\partial}{\partial k_{\rm ph}}+H^2k_{\rm ph}^2\f{\partial^2}{\partial k_{\rm ph}^2}\bigg)\bigg]\big\vert {f}^{\rm eq}(k_{\rm ph})\big\vert^2\,.\label{eq:PDeq}}

For the Bunch-Davies equilibrium state 
one then finds the ultraviolet expansions, now for simplicity in $n=4$ dimensions
\ea{
\tilde{\rho}^{\rm eq}({k}_{\rm ph})&=\f{k_{\rm ph}}{2}+\frac{H^2(1-6 \xi)+m^2}{4k_{\rm ph}}
+\frac{2H^2m^2(1-6 \xi)-m^4}{16k_{\rm ph}^3}+O(k_{\rm ph}^{-5}),\\
\tilde{p}^{\rm eq}({k}_{\rm ph})&=\frac{k_{\rm ph}}{6}-\frac{H^2(1-6 \xi)+m^2}{12k_{\rm ph}}
-\frac{2H^2m^2(1-6 \xi)-m^4}{16k_{\rm ph}^3}+O(k_{\rm ph}^{-5}).
}

A crude way to regularise the ultraviolet divergence is to introducing an ultraviolet cutoff at fixed physical momentum $\Lambda$.
One would then find that the ``bare'' energy density and pressure have the ultraviolet divergences
\ea{
\rho^{\rm eq}_{\Lambda}&\equiv\int^\Lambda\frac{d^3\mathbf{k}_{\rm ph}}{(2\pi)^3}\tilde{\rho}^{\rm eq}({k}_{\rm ph})
\nonumber \\&=\frac{1}{16\pi^2}\left[\Lambda^4+(H^2(1-6 \xi)+m^2)\Lambda^2+\left(H^2m^2(1-6 \xi)-\frac{1}{2}m^4\right)\log\Lambda\right]
+\mbox{finite}\label{eq:rholambda}
,\\
p^{\rm eq}_{\Lambda}&\equiv\int^\Lambda\frac{d^3\mathbf{k}_{\rm ph}}{(2\pi)^3}\tilde{p}^{\rm eq}({k}_{\rm ph})
\nonumber \\&=\frac{1}{16\pi^2}\left[\frac{\Lambda^4}{3}-\frac{1}{3}(H^2(1-6 \xi)+m^2)\Lambda^2-\left(H^2m^2(1-6 \xi)-\frac{1}{2}m^4\right)\log\Lambda\right]+\mbox{finite}.\label{eq:plambda}}

The counterterms $\delta\rho$ and $\delta p$ should be chosen in such a way that they cancel the divergences (\ref{eq:rholambda}), (\ref{eq:plambda}).
In principle, they should be formed from the covariant tensors obtained by varying a local action \cite{Markkanen:2013nwa}. In de Sitter space covariant conservation of the counter terms, as expressed in equation (\ref{equ:modeencons2}) and (\ref{equ:modeencons3}),
would then imply
$\delta \rho+\delta p=0$, so that effectively they correspond to a 
renormalisation of the cosmological constant or vacuum energy. 

However, we can see easily from Eqs.~(\ref{eq:rholambda}) and (\ref{eq:plambda}) that counterterms satisfying $\delta \rho+\delta p=0$ would not be sufficient in this case because $\rho^{\rm eq}_{\Lambda}+p^{\rm eq}_{\Lambda}$ is non-zero and divergent. It follows from the relation (\ref{equ:modeencons}) and the mode-by-mode energy conservation equation (\ref{equ:modeencons2}) that in equilibrium,
\ee{k_{\rm ph}^2\left(
\tilde{\rho}^{\rm eq}({k}_{\rm ph})+\tilde{p}^{\rm eq}({k}_{\rm ph})
\right)
=\frac{1}{3}\frac{d}{dk_{\rm ph}}k_{\rm ph}^3\tilde{\rho}^{\rm eq}({k}_{\rm ph}).\label{equ:econskph}}
Therefore this divergence is given purely by the boundary terms
\ea{
\rho^{\rm eq}_{\Lambda}+p^{\rm eq}_{\Lambda}
&=
\int_0^\Lambda \frac{dk_{\rm ph} k_{\rm ph}^2}{2\pi^2}
\left(
\tilde{\rho}^{\rm eq}({k}_{\rm ph})+\tilde{p}^{\rm eq}({k}_{\rm ph})
\right)
=\int_0^\Lambda \frac{dk_{\rm ph}}{6\pi^2}\frac{d}{dk_{\rm ph}}k_{\rm ph}^3\tilde{\rho}^{\rm eq}({k}_{\rm ph})
=\frac{\Lambda^3\tilde{\rho}^{\rm eq}({k}_{\rm ph})}{12\pi^2}
\nonumber\\
&=\frac{1}{12\pi^2}\left[
\Lambda^4+\frac{H^2(1-6\xi)+m^2}{2}\Lambda^2+\frac{2H^2m^2(1-6\xi)-m^4}{8}
\right]+O(\Lambda^{-2}\label{eq:ropp}).
}
Indeed, the non-vanishing of (\ref{eq:ropp}) can be traced to the fact that a cut-off violates general covariance, an issue which has been known already for a long time \cite{Zel'dovich:1968zz} and more recently discussed in \cite{Akhmedov:2002ts,Maggiore:2010wr2,Martin:2012bt}. There are three obvious ways of overcoming this problem. 

The first approach is to choose a regularization that does not break covariance.
Dimensional regularization is a prime example of a covariant regularization prescription where the generated divergences can be regulated with counter terms satisfying $\delta \rho+\delta p=0$, as discussed in \cite{Akhmedov:2002ts,Martin:2012bt}. This feature is present also when using the Pauli-Villars regularization \cite{Zel'dovich:1968zz}. 

Another approach would be to choose counter terms that cancel any non-covariant contributions (\ref{eq:ropp}) completely \cite{Maggiore:2010wr2}. In this approach the counter term contribution and the regulated bare energy-momentum tensor would both separately be covariance violating, but the renormalized result would be covariant. This provides a potential calculational advantage because the calculations can be carried out in four dimensions.
In de Sitter space this especially useful since finding analytical formulae for $n$-dimensional integrals containing the mode solution (\ref{equ:feq}) is a highly non-trivial problem.

A third approach would be to use a method where each Fourier component has a counter term so that no regulator is needed. An example of the third approach,
would be the adiabatic prescription\footnote{For details of this technique we refer the reader to \cite{Birrell:1982ix,Parker:2009uva}} of \cite{Parker:1974qw,Parker:1974qw1,Bunch:1980vc}, which is one of the most popular renormalization techniques in curved backgrounds and is also a covariance respecting technique.
\begin{figure}
\begin{center}
\hspace{2mm}\includegraphics[width=0.75\textwidth,trim={3cm 22cm  4cm  2cm },clip]{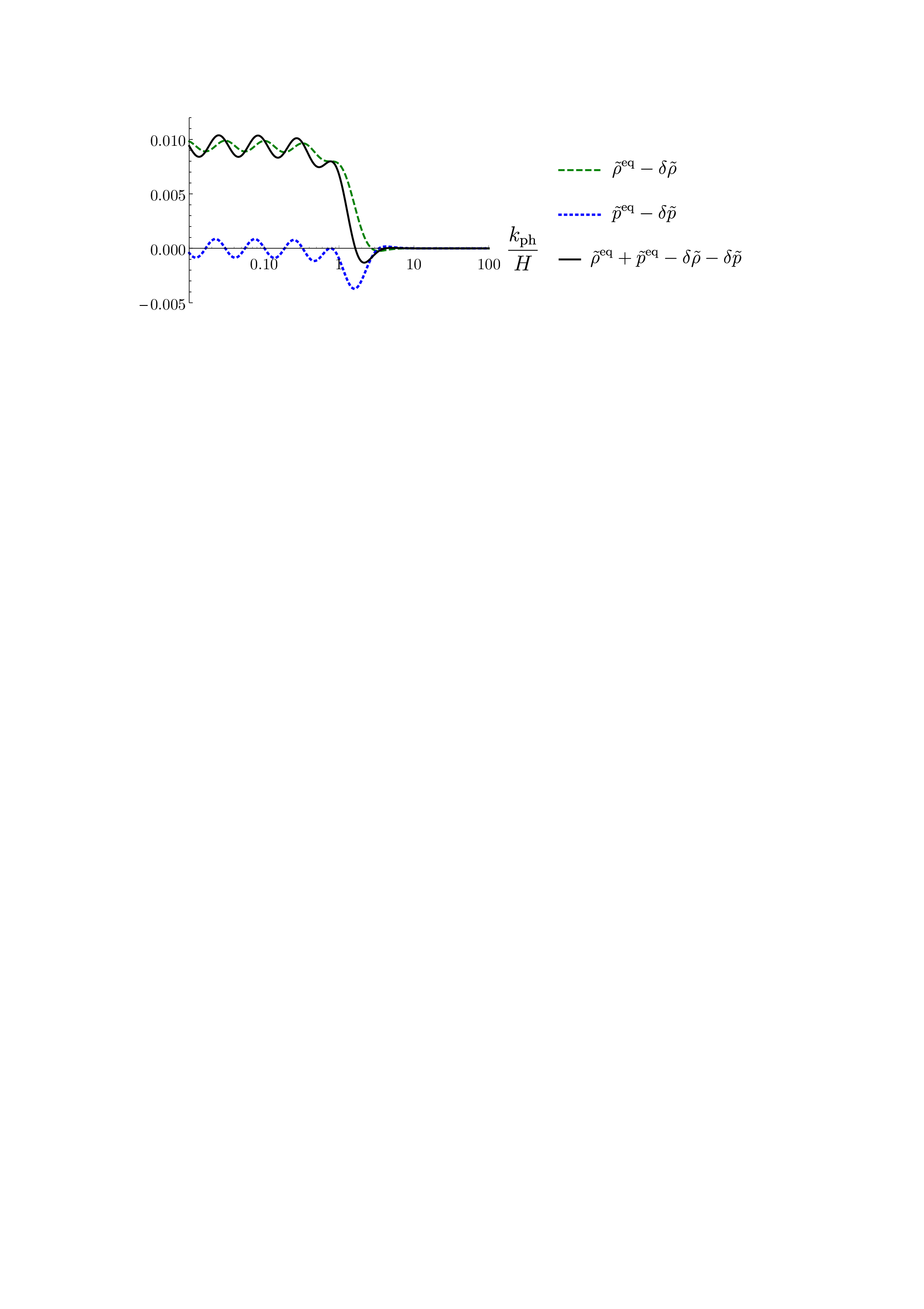}
\includegraphics[width=0.75\textwidth,trim={3cm 22cm  4cm  2cm },clip]{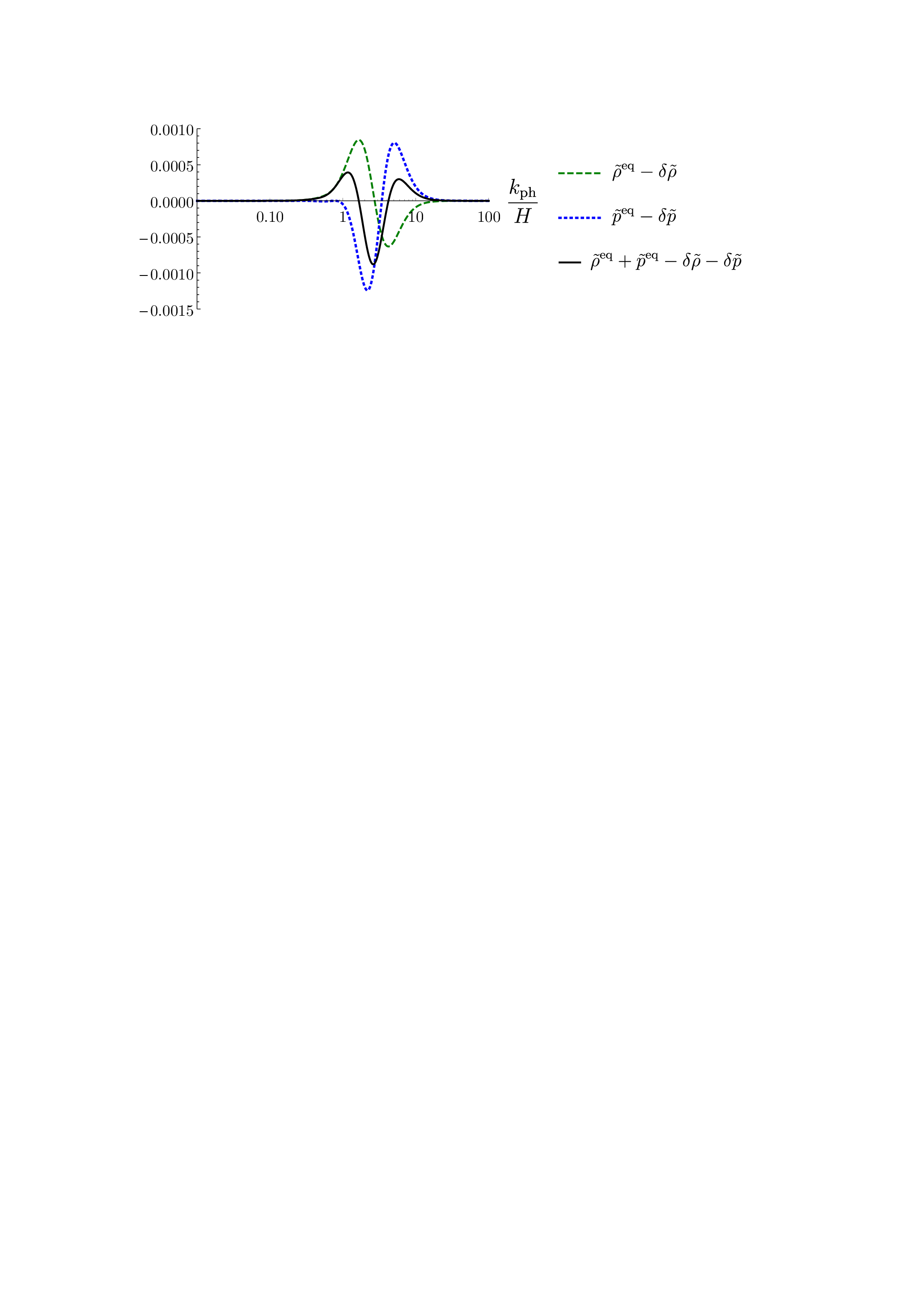}
\end{center}
\caption{\label{fig:modecontrib}}
Top: Contributions $\tilde{\rho}^{\rm eq}(k_{\rm ph})-\delta\tilde{\rho}(k_{\rm ph})$, $\tilde{p}^{\rm eq}(k_{\rm ph})-\delta\tilde{p}(k_{\rm ph})$ and $\tilde{\rho}^{\rm eq}(k_{\rm ph})+\tilde{p}^{\rm eq}(k_{\rm ph})-\delta\tilde{\rho}(k_{\rm ph})-\delta\tilde{p}(k_{\rm ph})$, from individual modes to the energy density $\rho$ (green dashed), pressure $p$ (blue dotted) and their sum $\rho+p$ (black solid), respectively, in the equilibrium Bunch-Davies state for $m/H=3$. \\
Bottom: The same mode contributions multiplied by $k^3/2\pi^2$ so that the overall quantity is given by the area under the curve, which integrates to zero for the sum $\tilde{\rho}^{\rm eq}(k_{\rm ph})+\tilde{p}^{\rm eq}(k_{\rm ph})-\delta\tilde{\rho}(k_{\rm ph})-\delta\tilde{p}(k_{\rm ph})$. 
\end{figure}
The reason no separate regularization is needed is that the adiabatic counterterms can be written as integrals over modes and therefore they can be combined under the same integral with the mode contributions so that, effectively, one is subtracting the energy density and pressure of the adiabatic vacuum mode--by--mode as
\ee{
\rho(t)-\delta\rho=\int\f{d^{3}\mathbf{k}_{\rm ph}}{(2\pi)^{3}}
\left(\tilde{\rho}_{\mathbf{k}_{\rm ph}}(t)-\delta\tilde{\rho}(k_{\rm ph})\right)\,;\qquad 
p(t)-\delta p=\int\f{d^{3}\mathbf{k}_{\rm ph}}{(2\pi)^{3}}
\left(\tilde{p}_{\mathbf{k}_{\rm ph}}(t)-\delta\tilde{p}(k_{\rm ph})\right)\,.\label{eq:renTmunu}} 
This makes the integrals convergent, and therefore no regularization is required. It also allows a direct investigation of individual mode contributions, although it should be borne in mind that they depend on the choice of the counterterms. In a sense adiabatic subtraction can be seen to combine the calculational simplicity of a cut-off and the strictly covariant nature of dimensional regularization.

For the equilibrium state with $m/H=3$ and $\xi=0$, these contributions are shown in Fig.~\ref{fig:modecontrib}. As the figure shows, the long-wavelength modes ($k_{\rm ph}\lesssim \sqrt{mH}$)
have positive renormalised energy density and their pressure is zero on average. This supports the interpretation that they carry a non-zero particle number. However, the modes in the ``particle production'' range $\sqrt{mH}\lesssim k_{\rm ph}\lesssim m^2/H$ have a negative pressure, 
which normal particle matter could not have.

Importantly, as one may explicitly show via Appendix~ \ref{sec:Ap}, the counter terms satisfy the the analogous mode-by-mode energy conservation condition to Eq.~(\ref{equ:econskph}), 
\ee{k_{\rm ph}^2\left(
\delta\rho(k_{\rm ph})+\delta{p}(k_{\rm ph})
\right)
=\frac{1}{3}\frac{d}{dk_{\rm ph}}k_{\rm ph}^3\delta{\rho}(k_{\rm ph}).\label{equ:econscounter}}
From this it follows that the renormalized energy density $\rho(t)$ and pressure $p(t)$ (\ref{eq:renTmunu}) satisfy the overall energy conservation equation (\ref{equ:modeencons2}).
As a special case, for the equilibrium state this implies that the sum of the renormalized energy and pressure density vanishes and hence that the equilibrium Bunch-Davies state has equation of state $w=-1$ for 
$p(t)=w\rho(t)$.
In spite of the apparent non-zero particle number, its energy density and pressure can therefore be fully absorbed into the vacuum energy. 

One can also explicitly verify from the expressions in Appendix~\ref{sec:Ap} that in dimensional regularisation the adiabatic counterterms integrate to zero, $\delta\rho+\delta p=0$.
This means that the two approaches agree up to a finite contribution to the vacuum energy, which can be absorbed into the bare cosmological constant.

However, even the adiabatic prescription is not ideal for practical calculations.
Even though the adiabatic counterterms give $\rho+p=0$ in equilibrium, both $\rho$ and $p$ are individually nonzero, and because of the arbitariness of the renormalisation scheme, their absolute value is not directly meaningful. Furthermore, as one can see from the expressions in Appendix~\ref{sec:Ap}, the adiabatic counter terms are rather lengthy and cumbersome, especially for a general $\xi$-parameter. Because of this in this work we will use, as far as we know, a novel renormalization technique where we consider the deviation of energy density and pressure from their equilibrium values,
\begin{eqnarray}
\rho(t)-\rho^{\rm eq}&=&\int\f{d^{3}\mathbf{k}_{\rm ph}}{(2\pi)^{3}}
\left(\tilde{\rho}_{\mathbf{k}_{\rm ph}}(t)-\tilde{\rho}^{\rm eq}(k_{\rm ph})\right),\nonumber\\
p(t)-p^{\rm eq}&=&\int\f{d^{3}\mathbf{k}_{\rm ph}}{(2\pi)^{3}}
\left(\tilde{p}_{\mathbf{k}_{\rm ph}}(t)-\tilde{p}^{\rm eq}(k_{\rm ph})\right).
\label{equ:rhominusreq}
\end{eqnarray}
These integrals are convergent and finite, and they are independent of the choice of the regularization scheme. Also, they are much more tractable analytically than the adiabatic prescription. 

To summarize, in this work we will render the integrals finite via (\ref{equ:rhominusreq}), and by interpreting 
$\tilde{\rho}^{\rm eq}(k_{\rm ph})$ and $\tilde{p}^{\rm eq}(k_{\rm ph})$ as renormalisation counterterms, this defines a natural and physically meaningful renormalisation scheme, which shares all the benefits of the adiabatic subtraction scheme without the need for the lengthy expressions resulting from an adiabatic expansion. We also point out that although the definition in (\ref{equ:rhominusreq}) results in a particular choice for the finite terms in the renormalized energy-density and pressure this is irrelevant for the combination $\rho(t)+p(t)$ due to the de Sitter symmetry of the equilibrium values and hence is not visible in the backreaction for $H$  studied in section \ref{sec:backreaction}.

\section{Equilibration}
\label{sec:att}

Let us now consider the case in which the initial state of the quantum field at some initial time $t=t_0$, at which $a(t_0)=1$, is not the de Sitter invariant Bunch-Davies state, but instead specified by some initial conditions $f_\mathbf{k}(t_0)$, $\dot{f}_\mathbf{k}(t_0)$. As the first step, we consider that the spacetime geometry is still de Sitter with a fixed Hubble rate $H$, but later in Section~\ref{sec:backreaction} we will discuss the backreaction from the quantum field to the metric. For simplicity, we also restrict ourselves to $n=4$ spacetime dimensions.

The general solution to the mode equation can be written as a linear combination in terms of the ``in'' mode
\ee{
f_\mathbf{k}(t)=C^{\rm in}_1(\mathbf{k})
f^{\rm in}_\mathbf{k}(t)+C^{\rm in}_2(\mathbf{k})
f^{\rm in*}_\mathbf{k}(t),
\label{equ:inicond}
}
where the coefficients $C^{\rm in}_1(\mathbf{k})$ and $C^{\rm in}_2(\mathbf{k})$ can be solved by using the normalization \ee{\dot{f}^{\,}_\mathbf{k}(t)f^*_\mathbf{k}(t)-f^{\,}_\mathbf{k}(t)\dot{f}^*_\mathbf{k}(t)=-i\,,} to give the initial conditions
\ea{
C^{\rm in}_1(\mathbf{k})&=-i\left(f_\mathbf{k}(t_0)\dot{f}^{\rm in\,*}_\mathbf{k}(t_0)-\dot{f}_\mathbf{k}(t_0)f_\mathbf{k}^{\rm in\,*}(t_0)\right)\,, \\
C^{\rm in}_2(\mathbf{k})&=i\left(f_\mathbf{k}(t_0)\dot{f}^{\rm in}_\mathbf{k}(t_0)-\dot{f}_\mathbf{k}(t_0)f_\mathbf{k}^{\rm in }(t_0)\right)
\,,\label{eq:incondi}}
which satisfy $|C^{\rm in}_1(\mathbf{k})|^2-|C^{\rm in}_2(\mathbf{k})|^2=1$. In precisely similar fashion one can also parametrise an initial condition with respect to the ``out'' solutions
\ee{f_\mathbf{k}(t)=C^{\rm out}_1(\mathbf{k})
f^{\rm out}_\mathbf{k}(t)+C^{\rm out}_2(\mathbf{k})
f^{\rm out*}_\mathbf{k}(t)\label{eq:incondo}\,,}
which can be related to the ``in'' basis via the Bogolubov coefficients (\ref{eq:B-E}) as
\ea{
C^{\rm out}_1(\mathbf{k})&=C^{\rm in}_1(\mathbf{k})\alpha_\mathbf{k}+C^{\rm in}_2(\mathbf{k})\beta^*_\mathbf{k}\,, \\
C^{\rm out}_2(\mathbf{k})&=C^{\rm in}_1(\mathbf{k})\beta_\mathbf{k}+C^{\rm in}_2(\mathbf{k})\alpha^*_\mathbf{k}
\,.}

Using Eqs.~(\ref{eq:ED1}) and (\ref{equ:inicond}), we can write for the energy density
\ea{
\rho_{\mathbf{k}}(t)
&=
\left(\f{1}{2}+|C^{\rm in/out}_2(\mathbf{k})|^2\right)\Bigg(
\left|\dot{f}^{\rm in/out}_{\mathbf{k}}(t)-\f{3H}{2}{f}^{\rm in/out}_{\mathbf{k}}(t)\right|^2
\nonumber\\&+
\left(\f{\mathbf{k}^2}{a^2}+m^2-2\xi\left(6H^2-3H\partial_t\right)\right)\left\vert {f}^{\rm in/out}_{\mathbf{k}}(t)\right\vert^2
\Bigg)
\nonumber\\
&+{\rm Re}
\Bigg\{
C^{\rm in/out}_1(\mathbf{k})C_2^{{\rm in/out *}}(\mathbf{k})
\Bigg[
\left(\dot{f}^{\rm in/out}_{\mathbf{k}}(t)-\f{3H}{2}{f}^{\rm in/out}_{\mathbf{k}}(t)\right)^2
\nonumber \\&+
\left(\f{\mathbf{k}^2}{a^2}+m^2-2\xi\left(6H^2-3H\partial_t\right)\right)\left( {f}^{\rm in/out}_{\mathbf{k}}(t)\right)^2
\Bigg]
\Bigg\}\,,\label{eq:ende}}
and pressure with Eq.~(\ref{eq:PD1})
\ea{
p_{\mathbf{k}}(t)
&=
\left(\f{1}{2}+|C^{\rm in/out}_2(\mathbf{k})|^2\right)\Bigg(
\left|\dot{f}^{\rm in/out}_{\mathbf{k}}(t)-\f{3H}{2}{f}^{\rm in/out}_{\mathbf{k}}(t)\right|^2
\nonumber\\ &-
\left(\f{\mathbf{k}^2}{3a^2}+m^2-2\xi\left(6H^2-4H\partial_t+\partial_t^2\right)\right)\left\vert {f}^{\rm in/out}_{\mathbf{k}}(t)\right\vert^2
\Bigg)
\nonumber\\
&+{\rm Re}
\Bigg\{
C_1^{\rm in/out}(\mathbf{k})C_2^{{\rm in/out *}}(\mathbf{k})
\Bigg[
\left(\dot{f}^{\rm in/out}_{\mathbf{k}}(t)-\f{3H}{2}{f}^{\rm in/out}_{\mathbf{k}}(t)\right)^2
\nonumber \\&-
\left(\f{\mathbf{k}^2}{3a^2}+m^2-2\xi\left(6H^2-4H\partial_t+\partial_t^2\right)
\right)\left( {f}^{\rm in/out}_{\mathbf{k}}(t)\right)^2\Bigg]
\Bigg\}.
\label{eq:pnde}}
In terms of physical momentum $\mathbf{k}_{\rm ph}=\mathbf{k}/a$ and choosing the ``in'' basis,
these can be written with the help of (\ref{eq:EDeq}) and (\ref{eq:PDeq}) as
\ea{
\tilde\rho_{\mathbf{k}_{\rm ph}}(t)
&=
\left(1+2|C^{\rm in}_2(a\mathbf{k}_{\rm ph})|^2\right)\tilde\rho^{\rm eq}({k}_{\rm ph})
\nonumber\\
&+{\rm Re}
\Bigg\{
C^{\rm in}_1(a\mathbf{k}_{\rm ph})C_2^{{\rm in *}}(a\mathbf{k}_{\rm ph})
\Bigg[
H^2\left(k_{\rm ph}{f}_{\rm eq}'(k_{\rm ph})+\f{3}{2}{f}_{\rm eq}(k_{\rm ph})\right)^2
\nonumber \\&+
\left(\mathbf{k}_{\rm ph}^2+m^2-2\xi\left(6H^2+3H^2k_{\rm ph}\frac{\partial}{\partial k_{\rm ph}}\right)\right)
{f}_{\rm eq}(k_{\rm ph})^2
\Bigg]
\Bigg\}\,,\label{eq:ende2}}
and

\ea{
\tilde{p}_{\mathbf{k}_{\rm ph}}(t)
&=
\left(1+2|C^{\rm in}_2(a\mathbf{k}_{\rm ph})|^2\right)\tilde p^{\rm eq}({k}_{\rm ph})
\nonumber\\
&+{\rm Re}
\Bigg\{
C_1^{\rm in}(a\mathbf{k}_{\rm ph})C_2^{{\rm in *}}(a\mathbf{k}_{\rm ph})
\Bigg[
H^2\left(k_{\rm ph}{f}_{\rm eq}'(k_{\rm ph})+\f{3}{2}{f}_{\rm eq}(k_{\rm ph})\right)^2\nonumber \\&-
\left(\f{\mathbf{k}_{\rm ph}^2}{3}+m^2-2\xi H^2\left(6+5k_{\rm ph} \frac{\partial}{\partial k_{\rm ph}} + k_{\rm ph}^2\frac{\partial^2}{\partial k_{\rm ph}^2}\right)
\right){f}_{\rm eq}(k_{\rm ph})^2\Bigg]
\Bigg\},
\label{eq:pnde2}}
where the only time dependence enters through the scale factor $a$. 

\begin{figure}
\begin{center}
\includegraphics[width=0.92\textwidth,trim={3.5cm 17cm  1.9cm  2.5cm },clip]{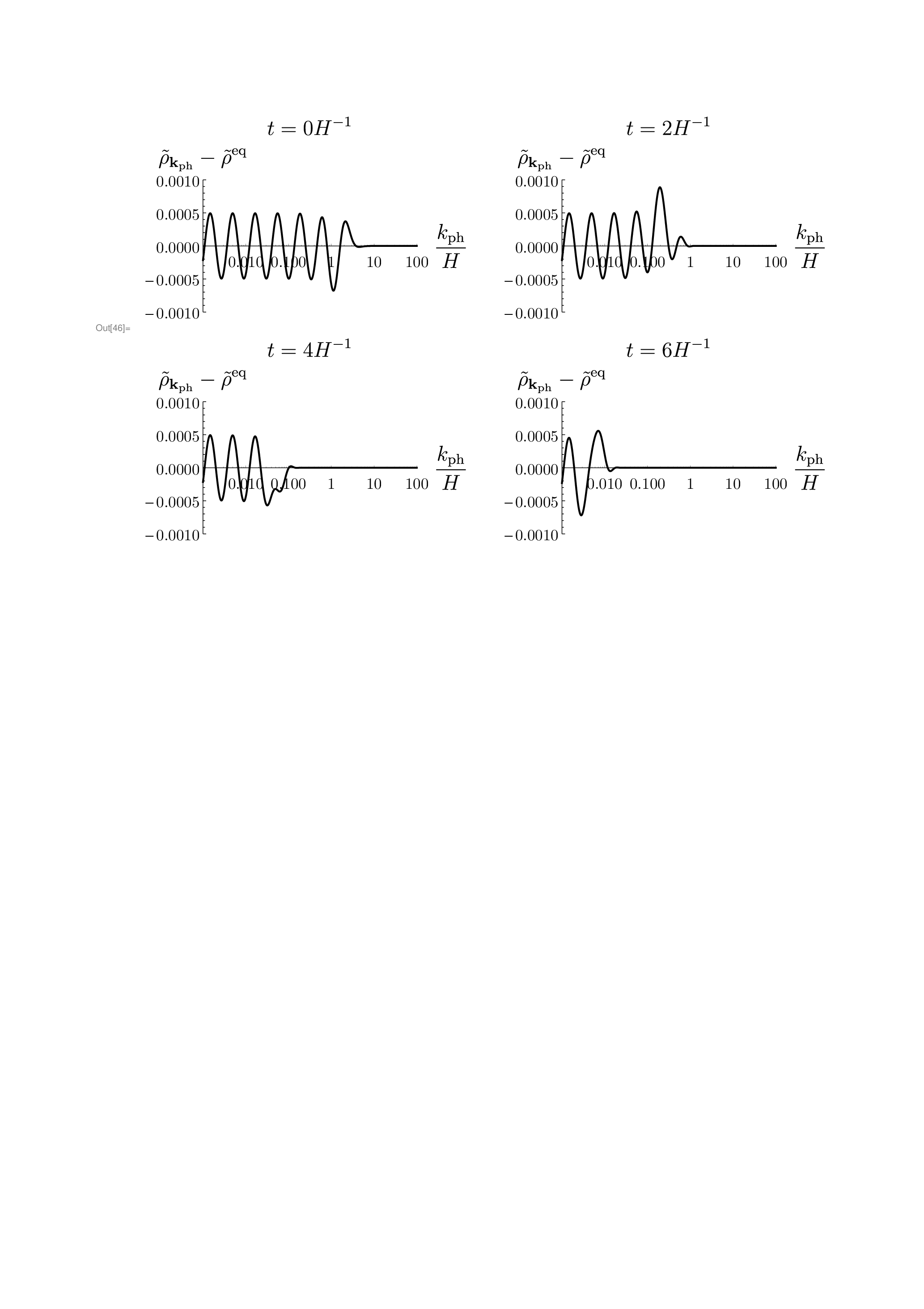}
\caption{
\label{fig:rhogrowth}
The energy density per mode $\tilde{\rho}_{\mathbf{k}_{\rm ph}}(t)-\tilde{\rho}^{\rm eq}(k_{\rm ph})$ as a function of the physical momentum $k_{\rm ph}$ plotted at different times for the adiabatic vacuum initial state with $m/H=3$ and $\xi=0$.
}
\end{center}
\end{figure}

\begin{figure}
\begin{center}
\includegraphics[width=7cm]{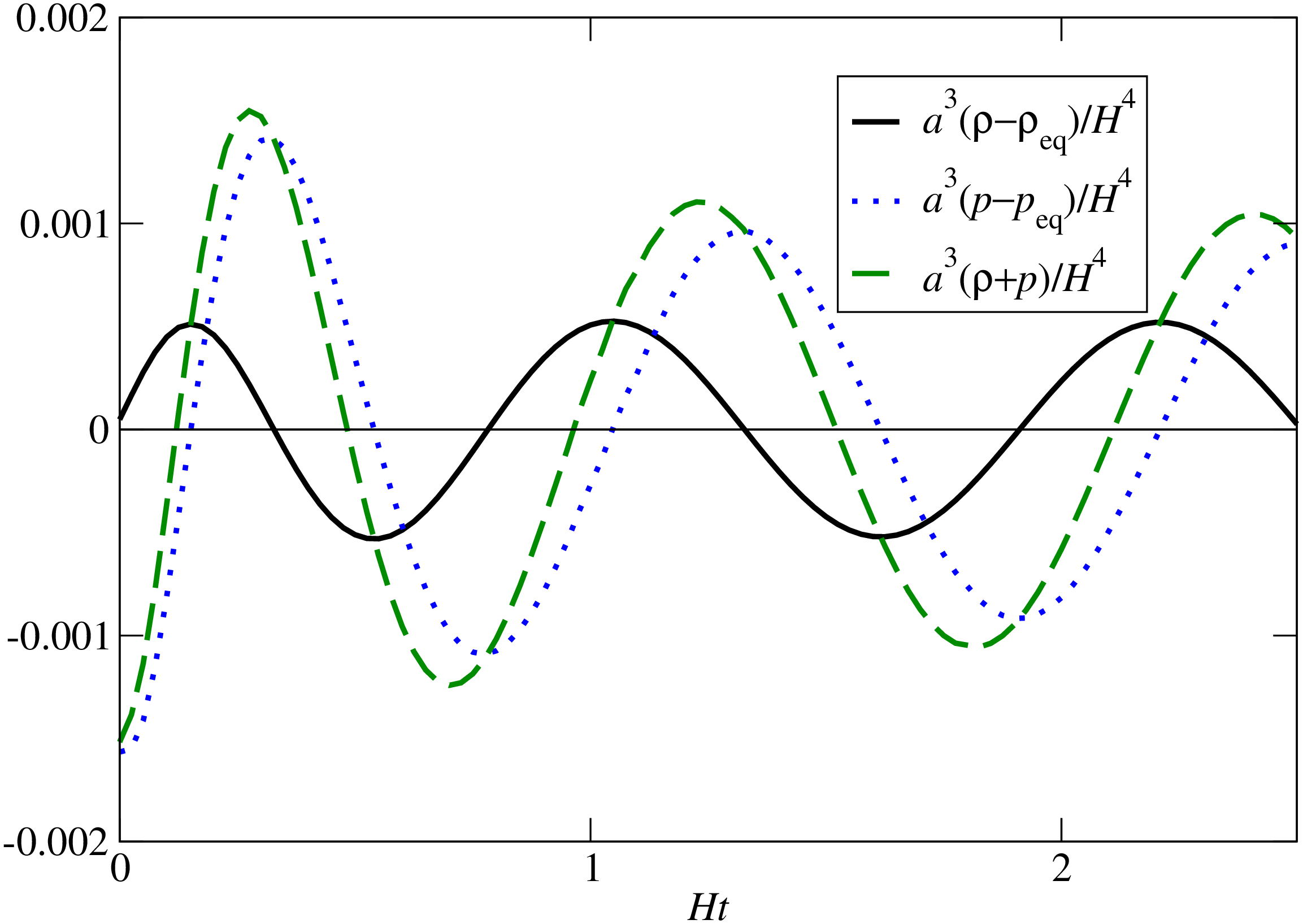}
\includegraphics[width=7cm]{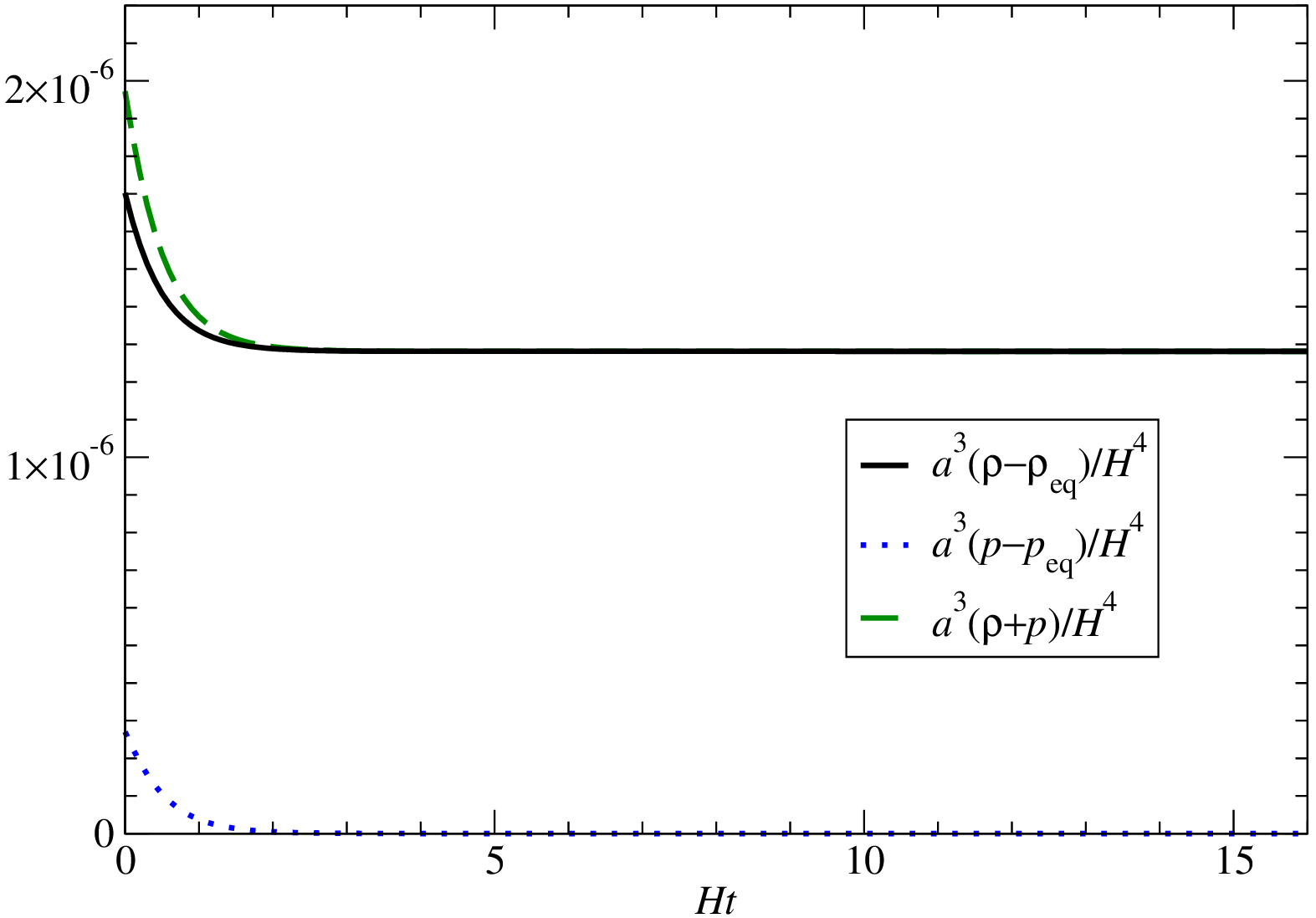}
\end{center}
\caption{\label{fig:rhoevol}
Left: Time evolution of the energy density and pressure starting from the adiabatic vacuum initial state for $m/H=3$.
For clarity, the curves show the deviation from the equilibrium value scaled by $a(t)^3$.
Right: The same quantities plotted without the oscillating second terms in Eqs.~(\ref{eq:ende}) and (\ref{eq:pnde}). This shows that if the oscillations are ignored, the field approaches a matter-like equation of state in a few e-foldings. The energy density of this matter-like component is much lower than the oscillating contribution.
}
\end{figure}

As an example, Figs.~\ref{fig:rhogrowth} and \ref{fig:rhoevol} show the time evolution of the energy density in momentum space for the case in which the field is initially in the fourth-order adiabatic vacuum state (\ref{equ:fad4}) at time $t=0$ and where we subtract the equilibrium contribution as in (\ref{equ:rhominusreq}). Although this state is sometimes considered to be an approximate vacuum state, it actually has a higher energy density than the equilibrium state. Starting from this state, the energy density oscillates around its equilibrium value with an amplitude that decreases as $a^{-3}$. 
In an interacting theory these oscillations would gradually lose coherence through particle scattering.

To investigate the late-time asymptotic behaviour, let us consider the initial condition where, at $t=0$, modes with momentum above a given cut-off $k_{\rm max}$ are initialized to the ``in'' vacuum to ensure the correct UV behaviour, and for modes below the cut-off we use an arbitrary initial condition. 

As discussed in Section~\ref{sec:br}, the ``out'' mode can be interpreted as the empty vacuum state for modes with $k_{\rm ph}\lesssim \sqrt{mH}$. Based on this notion we parametrize the initial condition 
with respect to the ``out'' modes as in (\ref{eq:incondo}).
Via (\ref{eq:ende}), this gives in the late time limit for the energy density
\ea{
\rho_{\mathbf{k}}(t)
&\underset{t\rightarrow\infty}{\longrightarrow}
\left(\f{1}{2}+|C^{\rm out}_2(\mathbf{k})|^2\right)\frac{m^2}{\gamma  H}
\nonumber\\
&+{\rm Re}
\bigg\{
C^{\rm out}_1(\mathbf{k})C_2^{{\rm out *}}(\mathbf{k})e^{-i \gamma H t}
\bigg[\frac{m^2}{2 \gamma  H}-\frac{\gamma  H}{2}+\frac{3}{2} i H (1-4 \xi )+
\frac{H (9-48 \xi )}{8 \gamma }
\bigg]
\bigg\}\,,\label{eq:ende1}}
and from (\ref{eq:pnde}) for the pressure
\ea{
p_{\mathbf{k}}(t)
&\underset{t\rightarrow\infty}{\longrightarrow}
{\rm Re}
\bigg\{
C^{\rm out}_1(\mathbf{k})C_2^{{\rm out *}}(\mathbf{k})e^{-i \gamma H t}
\nonumber \\ &\times\bigg[-\frac{m^2}{2 \gamma  H}-\frac{1}{2} \gamma  H (1-8 \xi )+2 i H \left(\frac{3}{4}-4 \xi \right)+\frac{H (9-48 \xi )}{8 \gamma }
\bigg]
\bigg\}\,.\label{eq:pnde1}}
If we have $m\gg H$ and $\xi\sim 1$, 
we can use (\ref{eq:ende1}) and (\ref{eq:pnde1}) to write
\ea{
{\rho}(t)-\rho^{\rm eq}~~&
\underset{t\rightarrow\infty}{\longrightarrow}~~
\frac{1}{a(t)^3}\int_0^{k<k_{\rm max}} \frac{d^3\mathbf{k}}{(2\pi)^3}\left[\left(|C^{\rm out}_2(\mathbf{k})|^2-|\beta_\mathbf{k}|^2\right)\frac{m^2}{\gamma H}\right],\nonumber\\
p(t)-p^{\rm eq}~~&
\underset{t\rightarrow\infty}{\longrightarrow}~~0,\label{equ:rhoasymp}}
where we have dropped the oscillating contributions as when $m\gg H$ on time scales of $\Delta t\sim 1/H$ they on average give a negligible contribution. 
Because $\rho^{\rm eq}+p^{\rm eq}=0$, the sum of the energy density and pressure at late times is
\ea{
{\rho}(t)+{p}(t)~~&
\underset{t\rightarrow\infty}{\longrightarrow}~~
\frac{1}{a(t)^3}\int_0^{k<k_{\rm max}} \frac{d^3\mathbf{k}}{(2\pi)^3}\left[\left(|C^{\rm out}_2(\mathbf{k})|^2-|\beta_\mathbf{k}|^2\right)\frac{m^2}{\gamma H}\right]
\,,\label{eq:lar}}
The integrals in Eqs.~(\ref{equ:rhoasymp}) and (\ref{eq:lar}) are time independent, and therefore the overall time dependence is $\propto a(t)^{-3}$. 
This means that the late time approach to the equilibrium state looks similar to classical matter, with the exception that the coefficient of $a^{-3}$ can be negative if $|C_2^{\rm out}|<|\beta_{\mathbf{k}}|$. In that case the apparent matter density is negative, corresponding to an overall equation of state with $w<-1$.

To conclude this section we note that the behaviour visible in (\ref{eq:lar}) and in Fig.~\ref{fig:rhoevol} where the equilibrium configuration is approached exponentially fast from non-de Sitter invariant initial conditions is a generic feature when the UV behaviour is close to that in Minkowski space. This can be understood from the simple observation that all physical modes are constantly redshifted and after a sufficient time scale, which for exponential expansion can be taken as $\sim 1/H$, only modes deep in the infrared can have any recollection of the initial condition and virtually the entire physical spectrum must be given in terms of the ``in'' or equilibrium solution (\ref{equ:feq}).

\section{Back Reaction}
\label{sec:backreaction}
In this section we address the problem of explicitly calculating the back reaction onto the Hubble rate in the semi-classical approximation where the expectation value of the energy momentum tensor calculated in a fixed de Sitter backround acts as the source in the Einstein equations. This is a good approximation when the deviation from equilibrium is small so that Einstein's equations can be linearised around the equilibrium de Sitter solution.

For a theory with a classical vacuum energy and pressure $\rho_{\rm vac}=-p_{\rm vac}$ and 
quantum field contributions $\rho(t)$ and $p(t)$, as defined in Eq.~(\ref{equ:Tmunu0}), 
we can write the semi-classical Einstein equations as
\ea{
\begin{cases}\phantom{-(}
3H^2M_{\rm pl}^2&= \rho(t)+\rho_{\rm vac}\\
 -(3H^2+2\dot{H})M_{\rm pl}^2 &= p(t)+p_{\rm vac} \end{cases}\,,\label{eq:def}}
where $M^2_{\rm pl}\equiv 1/(8\pi G)$. 
Summing the equations (\ref{eq:def}) allows us to write an evolution equation for $H$
\ee{2\dot{H}M_{\rm pl}^2=-\Big(\rho(t)+p(t)\Big)
\label{eq:Hev}\,.}

In the equilibrium state $\rho(t)=\rho^{\rm eq}$, and therefore the Hubble rate is
\begin{equation}
H_{\rm eq}^2=\frac{\rho^{\rm eq}+\rho_{\rm vac}}{3M_{\rm Pl}^2}.
\end{equation}
Since for the equilibrium solutions we have $\rho^{\rm eq}+p^{\rm eq}=0$ from the evolution equation (\ref{eq:Hev}) we get $\dot{H}=0$, in agreement with this.
There is no sign of a component behaving as a density of classical matter particles, since classical particles never have negative pressure and hence must give a non-zero contribution on the right hand side of (\ref{eq:Hev}). This is in spite there being two inequivalent mode solutions that can be interpreted as vacuous in the remote past and distant future as discussed in section \ref{sec:br}. Because of this we can argue that alone the existence of a non-trivial Bogoluybov transformation between vacua does not necessarily imply that there is a back reaction contribution with an interpretation as that of particles created from the vacuum. As far as the back reaction is concerned non-zero contributions result only when we start from non-equilibrium or non-de Sitter invariant initial conditions.

For small deviations from equilibrium, when the linear approximation is valid, we therefore have
\begin{equation}
H^2(t)=H_{\rm eq}^2+\frac{\rho(t)-\rho^{\rm eq}}{3M_{\rm Pl}^2},
\end{equation}
and therefore Eqs.~(\ref{equ:rhominusreq}) and (\ref{eq:EDeq})
determine the evolution of the Hubble rate.
Because $\rho(t)\rightarrow \rho^{\rm eq}$ as $t\rightarrow\infty$, the de Sitter state is stable against homogeneous perturbations of the field.

Using the expression (\ref{eq:lar}) valid in the late time limit and for $m\gg H$ we can probe the consequences of various initial conditions analytically. For completeness we use the cut-off $\Lambda=\sqrt{mH}$ motivated by our discussion in section \ref{sec:br} indicating that when the physical momentum is below this cut-off the ``out'' mode approximately serves as the vacuum and is the natural basis for the initial condition. The initial condition with $\vert C_2^{\rm out}(\mathbf{k})\vert=\vert\beta_\mathbf{k}\vert$ is simply the equilibrium choice $f^{\rm  }_\mathbf{k}=f^{\rm in}_\mathbf{k}$ for which the right hand side of (\ref{eq:Hev}) vanishes and so this choice acts as the border between a positive and negative source term. An important special case is found by choosing an initial condition with the smallest possible contribution, namely $\vert C_2^{\rm out}(\mathbf{k})\vert=0$ or $f^{\rm  }_\mathbf{k}=f^{\rm out}_\mathbf{k}$ for modes below the cut-off, which results in
\ee{2\dot{H}M_{\rm pl}^2=\int_0^{k<\sqrt{m H}} \frac{d^3\mathbf{k}}{(2\pi a)^3}|\beta_\mathbf{k}|^2\f{m^2}{\gamma H}=\f{1}{a^3}\f{m(mH)^{3/2}}{6\pi^2}e^{-2\pi m/H}\,,\label{eq:Hev2}}
where the last equality follows from (\ref{eq:B-E}) at the limit $m\gg H$ and with $\xi\sim\mathcal{O}(1)$. Apparently, modes initialized to a state that is approximately the vacuum of a late time observer result in a contribution on the right hand side of (\ref{eq:Hev2}), which behaves very closely to that of a particle density initially as pressureless non-relativistic matter with the Gibbons-Hawking de Sitter temperature\footnote{A thermal energy density at the non-relativistic limit with $T_{\rm dS}=H/(2\pi)$ is \ee{\rho_{\rm Th}(t)=\int \f{d^3\mathbf{k}}{(2\pi)^3}\f{\sqrt{\mathbf{k}^2+m^2}}{\exp\{\,\sqrt{\mathbf{k}^2+m^2}/T_{\rm dS}\}-1}\approx\f{m(mH)^{3/2}}{(2\pi)^3}{e^{-2\pi m/H}}}}. Importantly, the quantum contribution to (\ref{eq:Hev2}) implies $\dot{H}>0$ i.e. a period of super-acceleration, which means that in fact this contribution has an interpretation as exitation of initially empty states as opposed to the dilution of particles. This can be understood from the fact that the attractor solution ``in'' (\ref{eq:hank}) is occupied with respect to the ``out'' solution (\ref{eq:bes}) as implied by the Bogolubov relation (\ref{eq:B-E}). This result indicates that also in the non-interacting massive case quantum corrections may violate the weak energy condition such that $\rho+p<0$, as was famously shown for the massless interacting case in \cite{Onemli:2004mb,Onemli:2002hr}. In both scenarios, this can be seen to be due to the tendency of the scalar field to equilibrate in de Sitter space: starting from an energy-density smaller than the equilibrium state will result in a period where the energy density increases, which in turn leads to an increase in the Hubble rate. This period is rather brief as this effect is quickly exponentially suppressed.

For the initial condition of the fourth order adiabatic vacuum, we can see from Fig.~\ref{fig:rhoevol}\, that also for this non-zero initial condition we may have $\rho+p<0$ and conclude that the violation of the weak energy condition is a generic feature in de Sitter space, even without interactions.

\section{Conclusions}
\label{sec:con}
The exponentially expanding de Sitter space
has significant relevance for the early and late time Universe. In this work we have studied the evolution of a massive, non-interacting and non-minimaly coupled scalar field in such a spacetime. The Bunch-Davies vacuum state enjoys a special status in de Sitter space and we have shown here that the de Sitter invariance of this state is independent of the details of renormalization and can be understood as a manifestation of covariant conservation. Importantly, although the behaviour of the modes in the Bunch-Davies state can be interpreted to constantly go through a particle creation process as indicated by a nontrivial Bogolubov transformation this is not visible in the semi-classical backreaction: It bears no sign of a density from classical particles and implies strictly $w=-1$ for the equation of state.

We also studied the behaviour of the energy-momentum when starting from non-de Sitter invariant initial conditions. It is known that the concept of a particle is non-trivial in curved space and in our approach there is no need to give the definition of a particle. Instead, the evolution is fully determined by the initial condition for the field modes. By effectively using the Bunch-Davies vacuum mode contributions as renormalisation counterterms, 
we were able to investigate the time dependence of the energy-momentum tensor without the need for
ultraviolet regularisation. 
Via analytic and numerical examples, we saw that although it is possible to obtain a initial period where the equation of state is $w\neq-1$ all initial conditions approach the Bunch-Davies state as an equilibrium configuration on timescales of $\sim 1/H$. This is in full agreement with the conclusions of \cite{Habib:1999cs,Anderson:2000wx,Koivisto:2010pj}.

Finally we considered the semi-classical backreaction onto the Hubble rate. Due to the tendency of the scalar field to equilibrate in de Sitter space, we concluded that the expanding patch in the FLRW coordinates is a stable configuration under small perturbations of the initial conditions. As an important result from a fundamental point of view we showed that the weak energy condition can in principle be violated by quantum corrections i.e. that the equation of state can be $w<-1$, even though this effect is quickly exponentially suppressed. This was verified at the late time limit via an analytic calculation as well as a numerical analysis. Thus in support of the analysis of a massless self-interacting scalar field in \cite{Onemli:2004mb,Onemli:2002hr} we obtain that also for a massive non-interacting scalar field quantum corrections may lead to a period of superacceleration with $\dot{H}>0$.

\acknowledgments{TM and AR would like to thank Mark Hindmarsh, Malcolm Fairbairn and Gerasimos Rigopoulos for illuminating discussions. The research leading to these results has received funding from the European Research Council
under the European Union's Horizon 2020 program (ERC Grant Agreement no.648680). TM is supported by the Osk. Huttunen foundation, and AR by STFC grant ST/L00044X/1.}
\appendix
\section{Adiabatic counter terms in de Sitter}
\label{sec:Ap}
Here we give the explicit $n$-dimensional counter terms for the energy momentum of a non-interacting scalar field in de Sitter space and the fourth order adiabatic mode as defined by the adiabatic subtraction technique \cite{Parker:1974qw,Parker:1974qw1,Bunch:1980vc,Parker:2009uva,Birrell:1982ix}. 

For the adiabatic mode we can write the positive frequency mode as
\ee{
f_{\mathbf{k}}(t)=\f{\exp\left\{-i\int^t W\right\}}{\sqrt{2W}}\,,
\label{equ:fad4}
}
where
\ea{W^2&=\omega_0^2+\bigg[\frac{1}{4} n (4 (n-1) \xi -n+2)-\frac{3 m^2}{2 \omega_0 ^2}+\frac{5 m^4}{4 \omega_0 ^4}\bigg]H^2+\bigg[\frac{3 n (-4 \xi  n+n+4 \xi -2)}{8 \omega_0 ^2}\nonumber\\&+\frac{m^2 (2 n (4 (n-1) \xi -n+2)+15)}{2 \omega_0 ^4}+\frac{m^4 (5 n (-4 \xi  n+n+4 \xi -2)-249)}{8 \omega_0 ^6}\nonumber\\&+\frac{81 m^6}{2 \omega_0 ^8}-\frac{135 m^8}{8 \omega_0 ^{10}}\bigg]H^4\,,}
and where $\omega_0^2=\mathbf{k}^2/a^2+m^2$. Using (\ref{equ:fad4}) for calculating the energy-momentum up to terms $\mathcal{O}(H^4)$  gives
\ea{\delta T_{00}&=\int\f{d^{n-1}\mathbf{k}}{(2\pi a)^{n-1}}\bigg\{\f{\omega_0}{2}+\bigg[-\frac{n^2 \xi }{4 \omega_0 }+\frac{n^2}{16 \omega_0 }+\frac{3
   n \xi }{4 \omega_0}-\frac{n}{4 \omega_0 }-\frac{\xi }{2 \omega_0 }+\frac{1}{4 \omega_0 }-\frac{m^2 n \xi }{2 \omega_0 ^3}+\frac{m^2 n}{8 \omega_0 ^3}\nonumber\\&+\frac{m^2 \xi }{2 \omega_0 ^3}-\frac{m^2}{4 \omega_0 ^3}+\frac{m^4}{16 \omega_0 ^5}\bigg]H^2+\bigg[\frac{3 \xi ^2 n^4}{16 \omega_0 ^3}-\frac{3 \xi  n^4}{32 \omega_0 ^3}+\frac{3 n^4}{256 \omega_0 ^3}-\frac{9 \xi ^2 n^3}{8 \omega_0 ^3}+\frac{21 \xi  n^3}{32 \omega_0 ^3}-\frac{3 n^3}{32 \omega_0
   ^3}\nonumber\\&+\frac{27 \xi ^2 n^2}{16 \omega_0 ^3}-\frac{21 \xi  n^2}{16 \omega_0 ^3}+\frac{15 n^2}{64 \omega_0 ^3}-\frac{3 \xi ^2 n}{4 \omega_0 ^3}+\frac{3 \xi  n}{4 \omega_0 ^3}-\frac{3 n}{16 \omega_0 ^3}+\frac{3 m^2 \xi ^2 n^3}{4 \omega_0 ^5}-\frac{3 m^2 \xi  n^3}{8 \omega_0 ^5}+\frac{3 m^2 n^3}{64 \omega_0 ^5}\nonumber\\&-\frac{3 m^2 \xi ^2 n^2}{2 \omega_0 ^5}+\frac{7 m^2 \xi  n^2}{8 \omega_0 ^5}-\frac{m^2 n^2}{8
   \omega_0 ^5}+\frac{3 m^2 \xi ^2 n}{4 \omega_0 ^5}+\frac{11 m^2 \xi  n}{8 \omega_0 ^5}-\frac{13 m^2 n}{32 \omega_0 ^5}-\frac{15 m^2 \xi }{8 \omega_0 ^5}+\frac{15 m^2}{16 \omega_0 ^5}\nonumber\\&+\frac{5 m^4 n^2 \xi }{32 \omega_0 ^7}-\frac{5 m^4 n^2}{128 \omega_0 ^7}-\frac{135 m^4 n \xi }{32 \omega_0 ^7}+\frac{35 m^4 n}{32 \omega_0 ^7}+\frac{65 m^4 \xi }{16 \omega_0 ^7}-\frac{145 m^4}{64 \omega_0
   ^7}+\frac{35 m^6 n \xi }{16 \omega_0 ^9}-\frac{35 m^6 n}{64 \omega_0 ^9}\nonumber\\&-\frac{35 m^6 \xi }{16 \omega_0 ^9}+\frac{7 m^6}{4 \omega_0 ^9}-\frac{105 m^8}{256 \omega_0 ^{11}}\bigg]H^4\bigg\} \label{eq:ad1}\\ \nonumber
   &\equiv\int\f{d^{n-1}\mathbf{k}_{\rm ph}}{(2\pi)^{n-1}}\delta\tilde{\rho}(k_{\rm ph})}
and  
\ea{\delta T_{ii}/a^2&=\int\f{d^{n-1}\mathbf{k}}{(2\pi a)^{n-1}}\f{\omega_0^2-m^2}{n-1}\bigg\{\f{1}{2\omega_0}+\bigg[\frac{\xi  n^2}{4 \omega_0 ^3}-\frac{n^2}{16 \omega_0 ^3}-\frac{3 \xi  n}{4 \omega_0 ^3}+\frac{n}{4 \omega_0 ^3}+\frac{\xi }{2 \omega_0 ^3}-\frac{1}{4 \omega_0 ^3}\nonumber\\&+\frac{3 m^2 n \xi }{2 \omega_0 ^5}-\frac{3 m^2 n}{8 \omega_0 ^5}-\frac{3 m^2 \xi }{2 \omega_0 ^5}+\frac{3 m^2}{4 \omega_0 ^5}-\frac{5 m^4}{16 \omega_0 ^7}\bigg]H^2+\bigg[-\frac{9 \xi ^2 n^4}{16 \omega_0 ^5}+\frac{9 \xi  n^4}{32 \omega_0 ^5}-\frac{9 n^4}{256 \omega_0 ^5}\nonumber\\&+\frac{27 \xi ^2 n^3}{8 \omega_0 ^5}-\frac{63 \xi  n^3}{32 \omega_0 ^5}+\frac{9 n^3}{32 \omega_0
   ^5}-\frac{81 \xi ^2 n^2}{16 \omega_0 ^5}+\frac{63 \xi  n^2}{16 \omega_0 ^5}-\frac{45 n^2}{64 \omega_0 ^5}+\frac{9 \xi ^2 n}{4 \omega_0 ^5}-\frac{9 \xi  n}{4 \omega_0 ^5}+\frac{9 n}{16 \omega_0 ^5}\nonumber\\&-\frac{15 m^2 \xi ^2 n^3}{4 \omega_0 ^7}+\frac{15 m^2 \xi  n^3}{8 \omega_0 ^7}-\frac{15 m^2 n^3}{64 \omega_0 ^7}+\frac{15 m^2 \xi ^2 n^2}{2 \omega_0 ^7}-\frac{35 m^2 \xi  n^2}{8 \omega_0 ^7}+\frac{5 m^2
   n^2}{8 \omega_0 ^7}\nonumber\\&-\frac{15 m^2 \xi ^2 n}{4 \omega_0 ^7}-\frac{55 m^2 \xi  n}{8 \omega_0 ^7}+\frac{65 m^2 n}{32 \omega_0 ^7}+\frac{75 m^2 \xi }{8 \omega_0 ^7}-\frac{75 m^2}{16 \omega_0 ^7}-\frac{35 m^4 n^2 \xi }{32 \omega_0 ^9}+\frac{35 m^4 n^2}{128 \omega_0 ^9}\nonumber\\&+\frac{945 m^4 n \xi }{32 \omega_0 ^9}-\frac{245 m^4 n}{32 \omega_0 ^9}-\frac{455 m^4 \xi }{16 \omega_0 ^9}+\frac{1015 m^4}{64
   \omega_0 ^9}-\frac{315 m^6 n \xi }{16 \omega_0 ^{11}}+\frac{315 m^6 n}{64 \omega_0 ^{11}}+\frac{315 m^6 \xi }{16 \omega_0 ^{11}}\nonumber\\&-\frac{63 m^6}{4 \omega_0 ^{11}}+\frac{1155 m^8}{256 \omega_0 ^{13}}\bigg]H^4\bigg\}\,,\label{eq:ad2}\\ \nonumber
   &\equiv \int\f{d^{n-1}\mathbf{k}_{\rm ph}}{(2\pi)^{n-1}}\delta\tilde{p}(k_{\rm ph})\,.}
When the dimensions are analytically continued to $n$, the relation $\delta T_{00}+\delta T_{ii}/a^2=0$ may be verified by using the standard formulae \cite{Peskin:1995ev}.

\end{document}